\documentclass[aps,pre,twocolumn,showpacs,showkeys,a4paper]{revtex4-1}
 \usepackage{graphicx}
 \usepackage{amsmath}
 \usepackage{amssymb}
 \usepackage{amscd}
 \usepackage{color}

\begin{document}

 \title{The effect of active fluctuations on the dynamics of particles, motors and hairpins}
 \author{Hans Vandebroek$^{(1)}$ and Carlo Vanderzande$^{(1,2)}$}
 \affiliation{(1) Faculty of Sciences, Hasselt University, 3590 Diepenbeek, Belgium\\
  (2) Instituut Theoretische Fysica, Katholieke Universiteit Leuven, 3001
  Heverlee, Belgium}

%\pacs{05-40.-a; 05-60.-k; 02.50.Ga}

 \begin{abstract} 
Inspired by recent experiments on the dynamics of particles and polymers in artificial cytoskeletons and in cells, we introduce a modified Langevin equation for a particle in an environment that is a viscoelastic medium and that is brought out of equilibrium by the action of active fluctuations caused by molecular motors. We show that within such a model, the motion of a free particle crosses over from superdiffusive to subdiffusive as observed for tracer particles in an {\it in vitro} cytoskeleton or in a cell. We investigate the dynamics of a particle confined by a harmonic potential as a simple model for the motion of the tethered head of kinesin-1. We find that the probability that the head is close to its binding site on the microtubule can be enhanced by a factor of two due to active forces. Finally, we study the dynamics of a particle in a double well potential as a model for the dynamics of DNA-hairpins. We show that the active forces effectively lower the potential barrier between the two minima and study the impact of this phenomenon on the zipping/unzipping rate.
 \end{abstract}

\maketitle

\section{Introduction}
At the nanoscale all physical, chemical and biological processes are influenced by the ever present thermal noise. This is especially true for all molecular and cellular biophysical processes. In theoretical models of these phenomena it is therefore common to use the theory of stochastic processes \cite{VanKampen07}. The motion of a particle in a cell, the dynamics of a molecular motor, the translocation of a biopolymer through a membrane, the folding of a protein, are but a few of the various processes that are described using Langevin, Fokker-Planck or discrete state master equations \cite{Phillips13}. In most of these studies one assumes that the dynamics under study takes place in a viscous solvent that is in equilibrium. The latter property is taken care of by assuming a fluctuation-dissipation relation between the strength of the viscous and random forces appearing in the Langevin equations. Such an approach however neglects two important properties of the cytosol: it is a dense environment that certainly is not just an ordinary viscous solvent, and it is out of equilibrium \cite{Luby00,Luby13}. The neglect of these aspects can be justified as a first approximation, or as relevant for experiments which are often performed in simple {\it in vitro} solvents where measurements can be made in a more precise and controlled way than in the complex environment of a cell. Yet, if one wants to understand how cells function, it is necessary to perform experiments {\it in vivo} and to interpret them using theories that take into account the real physical properties of the cytosol. In the present paper we present an extended Langevin model that takes into account the viscoelasticity and the out-of-equilibrium nature of the cytosol and see how it effects simple processes like the motion of a particle, the stepping of a molecular motor or the zipping of a DNA hairpin. 

Let us briefly discuss two important properties in which the cytosol differs from an ordinary solvent. 
Living cells are crowded. Extensive studies have tried to characterise the physical properties of this environment \cite{Hofling13} through, for example, passive rheological measurements in which one follows the motion of tracer particles. The mean squared displacement (MSD), $\sigma^2(t)$, of these particles is found to follow a power law $\sigma^2(t) \sim t^\alpha$, where the exponent $\alpha$ almost always is different from one. Several models \cite{Hofling13} have been proposed to explain this behaviour but there is now a growing concensus that it is due to viscoelasticity \cite{Weiss13}. Viscoelastic effects appear as soon as the size of the tracer particle becomes of the order of the mesh size of the dense polymer networks that are present in the cytosol. 

At the same time, the cell is a system out of equilibrium where various active processes (like the action of motor molecules (myosin) on the actin filaments of the cytoskeleton) lead to fluctuations which can modify the passive transport of internalised or endogenous particles and which can also modify the motion of chromosomal loci or other dynamical processes in biopolymers.

In an early study \cite{Caspi00} it was for example found that the mean squared displacement of a microsphere inside a living cell has a superdiffusive motion for times up to 10 seconds after which the MSD changed to a subdiffusive motion. The superdiffusive motion has to be attributed to nonequilibrium, active processes. Similar behaviour has by now been found in other experiments in cellular environments \cite{Gal12,Goldstein13,Reverey15} and for particles immersed in an {\it in vitro} model cytoskeleton that consists of actin filaments, crosslinkers and myosin motors \cite{Ses14,Stuhrmann12,Toyota11,Adar16}. 

Besides particles, also the dynamics of biopolymers like chromosomes and microtubuli in a cellular environment has been found to be different from that in an ordinary solvent. For example shape fluctuations of microtubuli in {\it in vitro} cytoskeletons were found to deviate from those expected for a semiflexible polymer in equilibrium \cite{Brangwynne08,Broedersz14}. The motion of chromosomal loci in simple organisms like bacteria and yeast was recently shown to depend on active forces \cite{Weber12,Javer12}. Weber {\it et al.} found that after addition of chemicals that inhibit ATP-synthesis, the diffusion constant of chromosomal loci decreased by $49\%$. In another study, superdiffusive motion of bacterial chromosomal loci has been observed \cite{Javer14}. Moreover, measurements of chromatin in eukaryotes show evidence for an important role played by ATP-dependent processes \cite{Zidovska13,Bronshtein15}. 

Recently the present authors \cite{Vandebroek15} studied the well known Rouse model of polymer dynamics \cite{Rouse53,Doi86} in a viscoelastic and active environment and found similarities between the motion of monomers in that model and the observed behaviour of chromosomal loci.

Less is known about the possible influence of active forces on other dynamical processes such as the stepping of molecular motors or the zipping of a DNA (or RNA) hairpin. 

Continuing the research presented in \cite{Vandebroek15} we here aim to further understand the influence of viscoelasticity and activity on some cellular processes through the study of the dynamics in a simplified model.
For this purpose, we study the motion of "particles" in various potentials and in an active viscoelastic environment. The "particle" can correspond with a real physical particle like a protein or a microsphere internalised in a cell, or could represent a slow variable or (reaction) coordinate used to describe for example the stepping of a motor domain of kinesin-1, or the zipping of a hairpin molecule. 

This paper is organised as follows. In section 2 we present and motivate our model. In section 3 we give an exact solution for a free particle and show that the behaviour of the MSD within our model is very similar to that observed in various experiments on the motion of internalised or endogenous particles in a cell or in an {\it in vitro} cytoskeleton. In section 4 we give an exact solution for a particle in a harmonic potential. Here we determine the transient and steady state behaviour of the position of the particle and use it to study the diffusion of the tethered motor domain of kinesin-1. In section 5 we study the motion in a double well potential where the "particle" has to be interpreted as a reaction coordinate. This case has to be studied numerically and the details of our algorithm are given in the supplementary information. We determine both the probability distribution of the position of the particle and the transition rate between the minima of the potential. This is as far we know the first study of the extension of Kramers' rate theory \cite{VanKampen07} to an active and viscoelastic environment. We apply our results to the zipping of hairpin molecules. Finally, in section 6 we present our concluding remarks.

\section{The model and its biophysical motivation}
As a starting point, we take the well known Langevin equation for the velocity $v(t)$ of a mesoscopic  particle of mass $m$ in a normal viscous fluid \cite{VanKampen07,Zwanzig01}
\begin{eqnarray}
m \frac{dv}{dt} = - \gamma v(t) - \frac{dV}{dx} + \xi_T(t)
\label{1}
\end{eqnarray}
Eq. (\ref{1}) is just Newton's equation of motion for a particle experiencing friction (with a friction coeffient $\gamma$) and subject both to a deterministic force from a potential $V(x)$ and a random thermal force $\xi_T(t)$. Most often $\xi_T(t)$ is taken to be a centered Gaussian random variable with an autocorrelation given by the fluctuation-dissipation theorem (FDT) $\langle \xi_T(t) \xi_T(t')\rangle = 2 \gamma k_B T \delta(t-t')$ where $\delta(t)$ is the Dirac delta function. 

%Eq. (\ref{1}) can be derived more rigourously from well established projection techniques in nonequilibrium statistical mechanics \cite{Zwanzig01} that integrate out the motion of fast variables to obtain an equation of motion of a slow variable, in this case the position of the tagged particle. 

%From (\ref{1}) it can be easily found that the VACF $\langle v(t) v(t') \rangle$ decays exponentially while the mean square displacement $\langle (x(t)-x(0) )^2 \rangle$ equals $2Dt$ where $D=k_BT/\gamma$ (Stokes-Einstein relation). Hence the motion is diffusive. 

In a viscoelastic environment \cite{Oswald14}, friction is history dependent and is described in terms of a kernel $K(t)$. For the simplest viscoelastic element, the Maxwell element \cite{Oswald14}, which consists of a spring and a dashpot, $K(t)$ is exponential. The Langevin equation (\ref{1}) is accordingly modified and becomes
\begin{eqnarray}
m \frac{dv}{dt} = - \gamma \int_{0}^t K(t-t') v(t') dt' - \frac{dV}{dx} + \xi_T(t)
\label{2}
\end{eqnarray}
The stochastic properties of the random force $\xi_T(t)$ are still determined by the FDT, which however in this case becomes $\langle \xi_T(t) \xi_T(t')\rangle =  \gamma k_B T K(t-t')$. The precise form of $K(t)$ depends on the rheological properties of the fluid. In the case that the viscoelasticity is due to polymers, which have a broad spectrum of relaxation times, $K(t)$ can very well be approximated by a power law on times scales that are smaller than that of the longest relaxation time. We will therefore take
\begin{eqnarray}
K(t) =(2-\alpha) (1-\alpha) t^{-\alpha}
\label{3}
\end{eqnarray}
with $0 \leq \alpha \leq 1$. 
With this choice of prefactor, the noise $\xi_T(t)$ becomes so called fractional Gaussian noise (fGn) \cite{Mandelbrot68}. In the limit $\alpha \to 1$ the viscous case (\ref{1}) is recovered. In this paper we will concentrate on the case $\alpha > 1/2$ which leads to the most interesting behaviour and which can also be simulated most efficiently with our numerical algorithm. Notice that (\ref{2}) still describes a system that evolves towards thermal equilibrium as guaranteed by the FDT. Only the transient behaviour of (\ref{1}) and (\ref{2}) is different. 

In this paper we will work in the overdamped limit ($m/\gamma \to 0$), as appropriate for the low Reynolds numbers relevant at the scales encountered in cellular biophysics.

Finally we need to model the active forces that put the system out of equilibrium. Here we use a model for active gels first introduced by Levine and MacKintosh \cite{MacKintosh08,Levine09}. It describes the {\it in vitro} cytoskeletons or the cytoplasm by a two-fluid model of a network of semiflexible polymers driven by molecular motors. It predicts an exponential correlation of active fluctuations, or, equivalently, a power spectrum that is constant at low frequencies $\omega$ and that decays as $\omega^{-2}$ at high frequencies. Such a form is indeed consistent with recent experiments on the fluctuations of carbon nanotubes inside cells \cite{Fakhri14}. More generally, one can expect that there is a typical time scale $\tau_A$ on which active processes are persistent, for example it could corresponds to the typical time that a myosin motor is attached to the actin filaments.

For these reasons we add to the Langevin equation (\ref{2}) an extra active noise term $\xi_A(t)$ that is a centered random variable with exponential correlation
\begin{eqnarray}
\langle \xi_A(t) \xi_A (t') \rangle = C \exp \left(- |t-t'| / \tau_A \right)
\label{4}
\end{eqnarray}
Here $C$ characterises the strength of the active processes. Since there is no friction term associated to the active forces and no related FDT, adding an active noise to (\ref{3}) makes the system a non-equilibrium one.

Experiments \cite{Ses14,Stuhrmann12,Toyota11} have shown that the displacement of a particle immersed in an artificial actomyosin network has a Gaussian distribution with exponential tails superimposed. However, in a very recent work \cite{Adar16} it has been shown that at low myosin concentrations the distribution is purely Gaussian. In this paper we will assume that $\xi_A(t)$ is a Gaussian random variable. Though we acknowledge that this is an approximation, it has the advantage that it leads to exactly solvable models which can then be used as benchmarks to investigate deviations from Gaussianity. The effect of non-Gaussian active forces will be investigated in future work. In this respect, we also mention recent work where the motion of a particle subject to thermal and non-Gaussian active noise was investigated \cite{BenIsaac15,Fodor15}. That work was limited to the viscous situation and also considered only the motion in an harmonic potential.

Putting everything together, the equation that we will study for various potentials $V(x)$ is the (generalized) Langevin equation
\begin{eqnarray}
\gamma \int_{0}^t K(t-t') v(t') dt' = - \frac{dV}{dx} + \xi_T(t) + \xi_A(t) \Theta(t)
\label{5}
\end{eqnarray}
where $\Theta(t)$ is a Heaviside function. Thus we assume that the system is in thermal equilibrium at $t=0$, at which time the active forces start to act. The resulting motion $x(t)$ ($t > 0$) of the particle can then be interpreted as a respons to the active forces. 

%We believe that this equation can be used as a first approximation to describe dynamical processes taking place in an active gel. Here we think in the first place to the {\it in vitro} cytoskeletons that consist of actin, myosin and a crosslinker, but as the results below will show our model can also be used to describe the motion of tracer particles within a cell.

%Finally, a note on orders of magnitude. We expect the strength of the active forces, $\sqrt{C}$, to be of the order of 1 to 10 pN, the typical forces exerted by molecular motors. In contrast, a first estimate for the strength of thermal forces can be made by assuming the medium to be viscous, but with a viscosity $\eta$ that is about 100 to 1000 larger than that of water. The strength of the thermal forces are then given by $\sqrt{\gamma k_B T}$ with $\gamma=6 \pi \eta a$ (Stokes relation). For a particle the size of a protein ($a=3$ nm) thermal forces are of the order $0.014$ pN, while for a microsphere, they are of the order $0.25$ pN. Hence the active forces are respectively $10^2 \sim 10^3$, $10 \sim 10^2$ stronger than the thermal ones. In some figures, we will work in units in which $\gamma k_B T=1$, implying that the relevant values of $C$ are in the range $10^2 \sim 10^6$. 

\section{Free particle}
The simplest case to study is that of a constant potential $V(x)$, i.e. the motion of a free particle in an active and viscoelastic environment. 

As a possible application we mention the movement of a particle inside a cell.
In several experiments it is observed, that the motion of such a particle is superdiffusive at early times, but switches to being subdiffusive after a time of the order of seconds. 
In \cite{Caspi00} it was for example found that the MSD of a microsphere inside a living cell has a superdiffusive motion $\sim t^{3/2}$ for times up to 10 seconds. On larger time scales, the particle had a subdiffusive motion $\sim t^{1/2}$. 
In a study of  breast-cancer cells \cite{Gal12,Goldstein13}  the role of molecular motors and filaments of the cytoskeleton in active transport was investigated. These authors tracked the motion of polystyrene particles in cells with different metastatic potential after chemical treatments that affect different cell components like myosin, actin and microtubules. 
While the precise value of the exponent depended on the treatment applied, in all cases it was found that the exponent $\alpha$ decreased from a value in the range $1.2 - 1.4$ to a significantly lower value of $0.8 - 1.0$ after approximately $3$ seconds. So also here there is a crossover between superdiffusive and subdiffusive motion. 
%In a recent work, Reverey et al. \cite{Reverey15} studied the motion of endogenous particles in the protist {\it A. castellanii}. The cytoplasm of this organism has been called supercrowded. In contrast to the earlier studies, the cells of {\it A. castellani} have a nonzero motility. After subtraction of this motion, the dynamics of vesicles and granules was still found to superdiffusive with an exponent $\sim 1.8$. For the times investigated (up to 10 seconds) no clear crossover to another exponent was observed. After treating the cells with latrunculin A or nocodazole, the exponent $\alpha$ decreased to value around $1.5 - 1.6$ in an early time regime. Moreover, a decrease of the exponent after a few seconds can be clearly observed in these cases, though the authors do not quote a value. 

Similar behaviour was also observed outside living matter for particles moving in actomyosin networks \cite{Stuhrmann12,Ses14}.

For a constant potential, the equation of motion (\ref{5}) becomes
\begin{eqnarray}
\gamma \int_{0}^t K(t-t') \dot{x}(t')dt' = \xi_T(t) + \xi_A(t) \Theta(t)
\label{11}
\end{eqnarray}
This linear equation can easily be solved by Laplace transformation techniques. The details will be given in the next section where we solve the more general problem of motion in an harmonic potential $V(x)=kx^2/2$. Taking the limit $k \to 0$ we find that for a free particle the MSD $\sigma^2(t)\equiv \langle (x(t)-x(0))^2 \rangle$ is given by: 
\begin{eqnarray}
\sigma^2(t) &=& \frac{2 D_\alpha t^\alpha}{\Gamma(\alpha+1)} \nonumber \\
&+& \frac{2 C \tau_A^{2\alpha}}{\Gamma^2(\alpha) \eta_\alpha^2} \int_0^{t/\tau_A} dy e^y y^{\alpha-1} \Gamma(\alpha;y, t/\tau_A)
\label{12}
\end{eqnarray}
Here $D_\alpha=k_B T/ \eta_\alpha$, $\eta_\alpha=\gamma \Gamma(3-\alpha)$ and 
\begin{eqnarray}
\Gamma(\alpha;x_1,x_2) = \int_{x_1}^{x_2} dx\ e^{-x} x^{\alpha-1}
\end{eqnarray}
is a difference of two incomplete gamma functions. The same MSD holds for the center of mass of a Rouse chain in a viscoelastic and active environment \cite{Vandebroek15}. 

\begin{figure}[h]
\centering
\includegraphics[width=9cm]{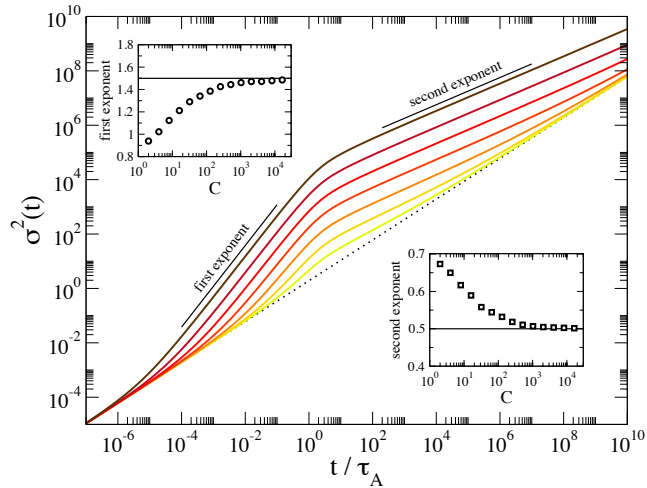}
\caption{Mean square displacement $\sigma^2(t)$ of a particle in an active viscoelastic environment with $\alpha=3/4$. The dotted line presents the behaviour in absence of active forces, while the various full lines correspond with $C=2^i$ with $i=2,4,6,8,10,12, 14$ (bottom to top). Time is measured in units of the persistence time $\tau_A$. The insets show the effective exponents for $t < \tau_A$ (upper left) and $t> \tau_A$ (lower right) as a function of the strength of the active forces $C$.}
\label{Fig1}
\end{figure}

Equation (\ref{12}) shows that in absence of active forces ($C=0$) the particle performs a subdiffusion with an exponent $\alpha$. To determine the behaviour in presence of active forces, we investigate the behaviour of the second term on the rhs of (\ref{12}). After an initial regime ($t\ll \tau_A$) in which this term can be neglected, it becomes proportional to $t^{2 \alpha}$ ($t < \tau_A$) while for  $t> \tau_A$ it evolves as $t^{2 \alpha-1}$ (See the supplementary material of \cite{Vandebroek15} for a detailed derivation of these results). The resulting behaviour of the MSD depends on the relative importance of the two terms in (\ref{12}) (see Fig.\ref{Fig1}). When $C \gg k_B T$ the second term dominates and the MSD changes from superdiffusive (with exponent $2 \alpha$) to subdiffusive 
(exponent $2 \alpha-1$). This scenario is consistent with the experimental results of \cite{Caspi00} where a drop in the exponent from $3/2$ to $1/2$ is observed. For that experiment we conclude that $\alpha=3/4$ and that the active forces are strong in comparison
with the thermal ones. Moreover from the experimentally determined time at which the exponent changes, one can conclude that $\tau_A \sim 10$ seconds. If however we decrease $C$, the two terms in (\ref{12}) can become comparable and hence one will observe an effective exponent that is between $2 \alpha$ and $\alpha$ for $t< \tau_A$ and between $\alpha$ and $2\alpha-1$ for $t> \tau_A$. This is shown in Fig.\ref{Fig1} for $\alpha=0.75$ and for various values of $C$. In the insets of Fig.\ref{Fig1} we show the dependence of the effective exponent on $C$ for $\alpha=3/4$ and $k_B T=1$. In these cases the drop in the effective exponent is less than $1$. For example, for $C=128$ the respective exponents are $1.39$ and $0.53$. We see that in these situations the observed exponents cannot be simply related to the viscoelastic properties of the medium. This could be the scenario behind the experiments on breast-cancer cells discussed above. 

We can thus conclude 
 that in the {\it nonequilibrium} situation there is no simple relation between the measured exponent of the MSD and the rheological property $\alpha$. In the relation between the two, an important role is played by the strength of the active fluctuations in comparison with the thermal ones.

\section{Particle in a harmonic potential}
We now turn to the case where the particle moves in a harmonic potential $V(x)=kx^2/2$. 

Before giving the solution of (\ref{5}) for this case, we discuss a possible application to molecular motors like kinesin-1. 
Kinesin-1 is a two-headed molecular motor whose two heads are connected with a linker that in turn is connected to a cargo-binding tail. Detailed mechanochemical models have been introduced that succesfully describe the velocity of the motor as a function of applied force, ATP-concentration, etc. It is less clear how to explain the large processivity of kinesin-1 and that of other motors in the kinesin family \cite{Hancock16}. A crucial role in these models is played by the mechanical properties of the neck linker. It is thought that after the binding of ATP to the front head, the free head has to perform a diffusive motion to the next site on the microtubule where it can then bind. This diffusive motion depends on the elastic properties of the neck linker, which in the simplest approximation is often modelled as a Hookean spring. A better approximation is to describe the linker as a wormlike chain (WLC) and use the latter's well known force-extension relation to determine the potential in which the motor head moves \cite{Kutys10}. Important properties are then the time it takes for the head to diffuse in this potential over a distance of 4.1 nm, i.e. the distance to the next binding site, and the probability $p$ that the head is within a small distance around that binding site. For example, in a model based on a Hookean spring, this probability was found to be rather small ($p=0.058$) \cite{Kutys10}. Below we will show that in active media, both the time to reach the binding site, and $p$ can be modified significantly.

In a harmonic potential, (\ref{5}) becomes
\begin{eqnarray}
\gamma \int_{0}^t K(t-t') v(t') dt' = - kx + \xi_T(t) + \xi_A(t) \Theta(t)
\label{14}
\end{eqnarray}
Since this is a linear equation, it can be solved using Laplace transform techniques. We will denote the Laplace transform of a function $g(t)$ by $\tilde{g}(s) = \int_0^\infty dt e^{-st} g(t) $. In this way, (\ref{14}) becomes
\begin{eqnarray}
\eta_\alpha s^{\alpha-1} ( s \tilde{x}(s) - x(0)) = - k \tilde{x}(s) + \tilde{\xi}_T (s) + \tilde{\xi}_A (s)
\label{15}
\end{eqnarray}
which gives
\begin{eqnarray}
\tilde{x}(s) &=& x(0) s^{-1} \left[1 + \frac{k s^{-\alpha}}{\eta_\alpha} \right]^{-1} \nonumber \\
&+& \frac{\tilde{\xi}_T (s) + \tilde{\xi}_A (s)}{\eta_\alpha} s^{-\alpha} \left[ 1 + \frac{k s^{-\alpha}}{\eta_\alpha} \right]^{-1}
\label{16}
\end{eqnarray}
The inverse Laplace transform can be done in terms of the Mittag-Leffler function $E_{\alpha,\beta}(z)$ which is an extension of the exponential function that appears often in studies of the motion in a viscoelastic medium with a power law kernel .
The Mittag-Leffler function \cite{Haubold11} is defined through its series expansion
\begin{eqnarray}
E_{\alpha,\beta}(z) = \sum_{i=0}^\infty \frac{z^i}{\Gamma(\alpha i + \beta)}
\label{17}
\end{eqnarray}
For $\alpha=\beta=1$ we recover the Taylor series of the exponential. From (\ref{17}), it can be shown that 
\begin{eqnarray}
\int_0^\infty dt\ e^{-st} t^{\beta-1} E_{\alpha,\beta}(a t^\alpha) = s^{-\beta} (1-a s^{-\alpha})^{-1}
\label{18}
\end{eqnarray}
Using (\ref{18}), one can invert (\ref{16}) and find the position of the particle 
\begin{eqnarray}
x(t) &=& x(0) E_{\alpha,1}\left(-(t/\tau)^\alpha\right) \nonumber \\ &+& \frac{1}{\eta_\alpha} \int_0^\infty dt' \left( \xi_T (t-t') + \xi_A(t-t') \right) t'^{\alpha-1} E_{\alpha,\alpha} \left( - (t'/\tau)^\alpha\right) \nonumber \\
\label{19}
\end{eqnarray}
where we have introduced the characteristic time
\begin{eqnarray}
\tau= \left(\eta_\alpha/k\right)^{1/\alpha}
\label{20}
\end{eqnarray}
Because (\ref{14}) is linear and the noises $\xi$ are Gaussian, also $x(t)$ is a Gaussian random variable. Hence it is sufficient to calculate its first two cumulants. It is immediately clear from (\ref{19}) that since we start from thermal equilibrium at $t=0$ (hence $\langle x(0) \rangle=0$), and since the noises are centered, that also $\langle x(t)Ê\rangle=0$ for all $t$. 

The calculation of the variance of $x(t)$ is more involved. Using the autocorrelations of the noises $\xi_T$ and $\xi_A$ and the fact that at $t=0$  the equilibrium equipartition theorem $\langle x^2(t)\rangle= k_BT /k$ holds (with $k_B$ Boltzmann's constant), one gets
\begin{align}
	&\langle x^2(t) \rangle = \frac{k_BT}{k} E_{\alpha,1}^2\Big(-\left(t/\tau\right)^{\alpha}\Big)  \nonumber \\ &+\frac{1}{\eta_{\alpha}^2} \int_0^t dt' \int_0^t dt'' \, \Big(\gamma k_BT\frac{(2-\alpha)(1-\alpha)}{|t'-t''|^{\alpha}} + Ce^{-|t'-t''|/\tau_A} \Big) \nonumber \\ &\times  t'^{\alpha-1}t''^{\alpha-1} E_{\alpha,\alpha}\Big(-\left(t'/\tau\right)^{\alpha}\Big)E_{\alpha,\alpha}\Big(-\left(t''/\tau\right)^{\alpha}\Big) \label{21}
\end{align}
This expression can be simplified greatly when we use a property of the Mittag-Leffler functions which is derived in the supplemental material. In this way, we obtain after some further manipulations
\begin{align}
	\langle x^2(t) \rangle = \frac{k_BT}{k} + \frac{C}{k^2} A_\alpha(t,\tau,\tau_A) \label{21bis}
\end{align}
Here we introduced the function $A_\alpha(t,\tau,\tau_A)$ which is given by
\begin{widetext}
\begin{align}
	A_\alpha(t,\tau,\tau_A) =  \int_0^{t/\tau} dx \int_0^{t/\tau} dy\  e^{-|x-y|\tau/\tau_A} x^{\alpha-1}y^{\alpha-1} E_{\alpha,\alpha}\Big(-x^{\alpha}\Big)E_{\alpha,\alpha}\Big(-y^{\alpha}\Big) .\label{22}
\end{align}
\end{widetext}
The first (second) term of (\ref{21bis}) is the contribution to the variance coming from thermal (active) fluctuations.

The probability density $P(x,t)$ that gives the chance to find the particle at time $t$ in the small interval between $x$ and $x+dx$ is then given by the Gaussian
\begin{align}
	P(x,t) = \frac{1}{\sqrt{2\pi \langle x^2(t)\rangle}} \exp \left( - \frac{x^2}{2\langle x^2(t)\rangle} \right)\label{23}
\end{align}
Notice that  {\it independently of $\alpha$}, $A_\alpha(t,\tau,\tau_A)$ is always strictly positive and therefore the density $P(x,t)$ always broadens after the active forces have been turned on. This broadening as a function of time is illustrated in Fig. 2 for $C=10^3$ and $\tau_A=1$ in a medium with $\alpha=3/4$.
\begin{figure}[h]
\centering
\includegraphics[width=9cm]{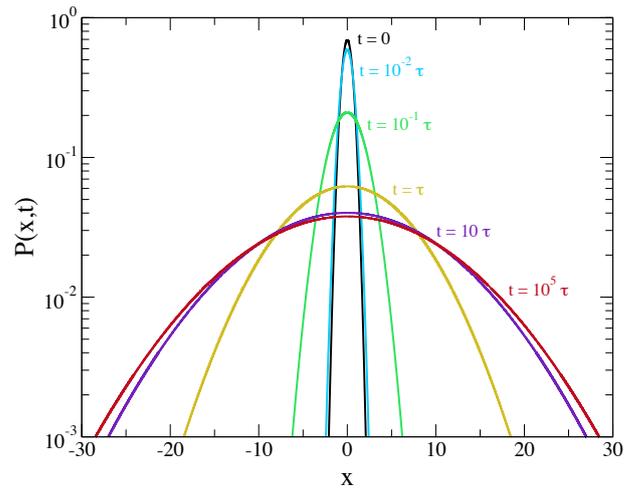}
\caption{Probability distribution $P(x,t)$ for a particle in a harmonic potential ($k=3$) after active forces with $C=10^3$ and $\tau_A=1$ are turned on at $t=0$.  At $t=0$ the distribution is that of thermal equilibrium with $k_B T=1$. The figure shows the broadening of the distribution as a function of time ($\tau=10^{-5}$)  in a medium with $\alpha=3/4$.}
\label{Fig2}
\end{figure}

For large times, the system will evolve to a new nonequilibrium steady state in which the variance reaches the value $\langle x^2\rangle_{ss} = \lim_{t \to \infty} \langle x^2(t) \rangle = k_B T/k + CA_\alpha(\infty,\tau,\tau_A)/k^2$.  From (\ref{22}) it can be seen that for $t \to \infty$, $A_\alpha(t,\tau,\tau_A)$ only depends on the ratio $\tau/\tau_A$. Hence we find that in the stationary state the variance is of the form

\begin{eqnarray}
\langle x^2 \rangle_{ss} = \frac{k_B T}{k}   + \frac{C}{k^2}\ f_\alpha(\tau/\tau_A)
\end{eqnarray}

In the limit $\tau/\tau_A \to 0$, the exponential function disappears from (\ref{22}) and since $\int_0^\infty dx\ x^{\alpha-1} E_{\alpha,\alpha} (-x^\alpha)=1$, it follows that $f_\alpha(\tau/\tau_A)
 \to 1$. In the reverse limit, $\tau/\tau_A \gg 1$, the exponential term in (\ref{22}) can be written in terms of a Dirac-delta function, from which it follows that $f_\alpha(\tau/\tau_A)$ goes  as
\begin{eqnarray}
f_\alpha(\tau/\tau_A) \to \frac{2\tau_A}{\tau} \int_0^\infty dx\  x^{2\alpha-2} E_{\alpha,\alpha}^2 (-x^\alpha)
\label{24}
\end{eqnarray}
In Fig.3 we plot $f_\alpha(\tau/\tau_A)$ for three values of $\alpha$. The important conclusion here is that the nonequilibrium steady state depends on $\alpha$, i.e. on the nature of the viscoelastic environment. This is in contrast to thermal equilibrium where there is no such dependence. 
\begin{figure}[h]
\centering
\includegraphics[width=9cm]{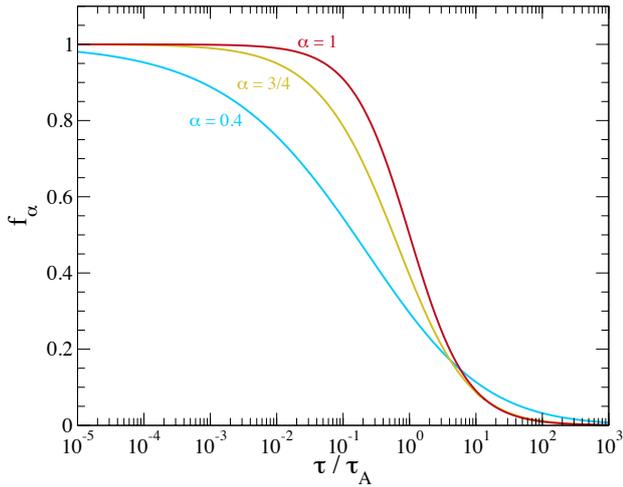}
\caption{The function $f_\alpha(\tau/\tau_A)$ (see text) for three values of $\alpha$.}
\label{Fig3}
\end{figure}

We now apply these results to the diffusive motion of the free head of a kinesin-1 motor, in which as explained above, the head linker is modelled as a Hookean spring. Here we follow \cite{Kutys10} and choose $k=1$ pN/nm. From (\ref{20}) it then follows that $\tau \approx 50 \mu$s, which is much shorter than the persistence time of the active forces, which is of the order of seconds (here as a rough estimate we took $\alpha=1$ and $\gamma=6\pi \eta a$ where for the cytosol we assumed $\eta$ to be thousand times that of water and have taken $a=3$ nm). Hence we are in the regime where $\tau/\tau_A \ll 1$, so that $\langle x^2 \rangle_{ss} = k_B T/k  + C/k^2$.  In the model of \cite{Kutys10} it is assumed that the head can attach to the microtubule when it is within one nm of the binding site at $4.1$ nm.
For the probability density (\ref{23}) in the stationary state, the probability $p$ that the head is in this range is given by
\begin{eqnarray}
p = \frac{1}{2} \left[ \mbox{erf} \left(\frac{5.1}{\sqrt{2 \langle x^2 \rangle_{ss}}}\right) - \mbox{erf} \left(\frac{3.1}{\sqrt{2 \langle x^2 \rangle_{ss}}}\right)\right]
\label{25}
\end{eqnarray}
where $\mbox{erf}(x)$ is the error function and all distances are expressed in nm.
In Fig. 4 we have plotted this probability as a function of the strength $\sqrt{C}$ of the active forces (in pN). We see that upon increasing $C$, $p$ increases from its value of $0.058$ in absence of active forces, reaches a maximum of $p=.118$ around $\sqrt{C} \approx 3.5$ pN and then decreases again. The value where $p$ reaches its maximum is indeed of the expected order for active forces in a cell. Hence, we conclude that active forces can double the probability that the free head is near its binding site. 
\begin{figure}[h]
\centering
\includegraphics[width=9cm]{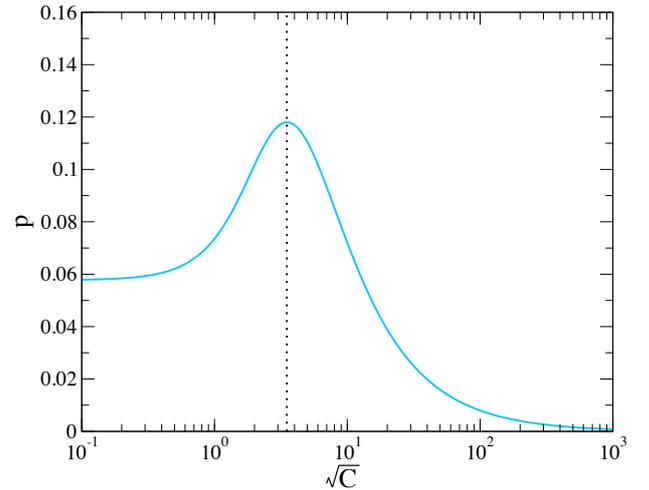}
\caption{Probability $p$ that the motor head is in a region of 1 nm around the binding site at $4.1$ nm as a function of the strength $\sqrt{C}$ of the active forces. This figure is for the regime $\tau/\tau_A \ll 1$ where $p$ only depends on $C$ and not on $\tau_A$ or $\alpha$.}
\label{Fig4}
\end{figure}

Next we look at the time dependence of the  MSD of the particle in the harmonic well. We have the relation
\begin{eqnarray}
\sigma^2(t) = \langle (x(t)-x(0))^2 \rangle = \langle x^2(t) \rangle + \langle x^2(0) \rangle - 2 \langle x(t) x(0) \rangle \nonumber \\
\label{26}
\end{eqnarray}
The only term that remains to be calculated is the third one, which can be easily found from (\ref{19}), the equipartition theorem at $t=0$ and the independence of the initial position and the random forces. Using also (\ref{21bis}) we find
\begin{eqnarray}
\sigma^2(t) = \frac{2 k_B T}{k} \left[1- E_{\alpha,1} \left(-(t/\tau)^\alpha\right)\right] + \frac{C}{k^2} A_\alpha(t,\tau,\tau_A) \nonumber \\
\label{27}
\end{eqnarray}

We observe that $\tau$ is the timescale on which the particle starts to experience the effects of the potential. Hence we expect that for $t \ll \tau$ the particle moves as a free particle \cite{remark}, whereas for $t \gg \tau$ it reaches its stationary state MSD. Therefore we predict that when $\tau_A \ll \tau$ the MSD will show the two regimes ($\sigma^2(t) \sim t^{2\alpha}$ followed by $\sigma^2(t) \sim t^{2\alpha-1}$) of the free particle before the MSD start to saturate. This is indeed the behaviour found if we plot (\ref{27}) for the case $\tau_A/\tau=10^{-5}$ (see Fig. 5). In the reverse case, which is the biophysically relevant one, $\tau \ll \tau_A$, there will be only the superdiffusive regime after which the MSD reaches its value at stationarity (see again Fig. 5). For intermediate values of $\tau_A/\tau$ the crossover to the stationary  value occurs either in the superdiffusive or subdiffusive regime. 
%In Fig.5 one can also notice that as soon as $\tau_A > \tau$, the precise value of $\tau_A$ has little influence on the MSD of the particle.

\begin{figure}[h]
\centering
\includegraphics[width=9cm]{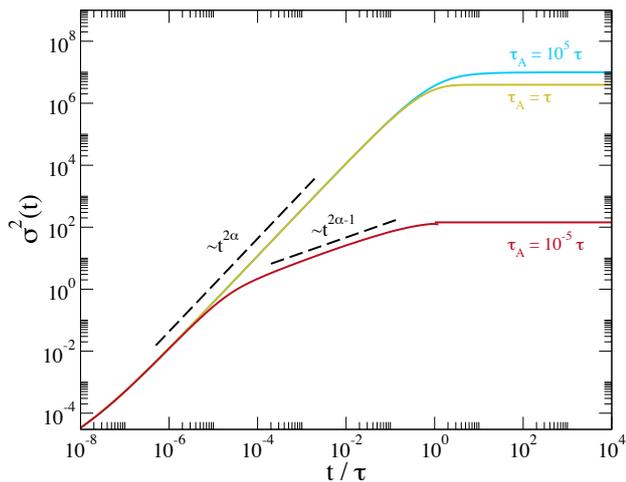}
\caption{Mean square distance travelled by a particle in a harmonic potential ($k=0.1$, $k_B T=1$) in a viscoelastic ($\alpha=3/4$) and active environment ($C=10^5$) for three different values of $\tau_A/\tau$.}
\label{Fig5}
\end{figure}

%In Fig. 5, the full lines are the result of a numerical evaluation of the exact expression (\ref{27}) while the dotted lines are the result of a simulation using the algorithm explained in the supplemental material. The perfect agreement between the two shows that the accuracy of the algorithm, which, in the next section will we used in a case where no exact solution is possible.

We have investigated the possible implication of the dynamics of the MSD on the motion of kinesin-1. For this we show in Fig. 6 the MSD for a particle in a viscous environment ($\alpha=1$) without active forces and in a viscoelastic environment ($\alpha=3/4$) in presence of active forces. We notice that in the latter situation the MSD is always somewhat larger. Given the uncertainty in the various parameters, we can conclude that the time to reach the binding site is of the same order in both situations. This suggests that active forces can be helpful in overcoming the expected slowing down of a molecular motor in a viscoelastic environment and thus may provide the answer why a motor in a cell moves with almost the same velocity as in an {\it in vitro} essay \cite{Cai07,Hill04}. 

\begin{figure}[h]
\centering
\includegraphics[width=9cm]{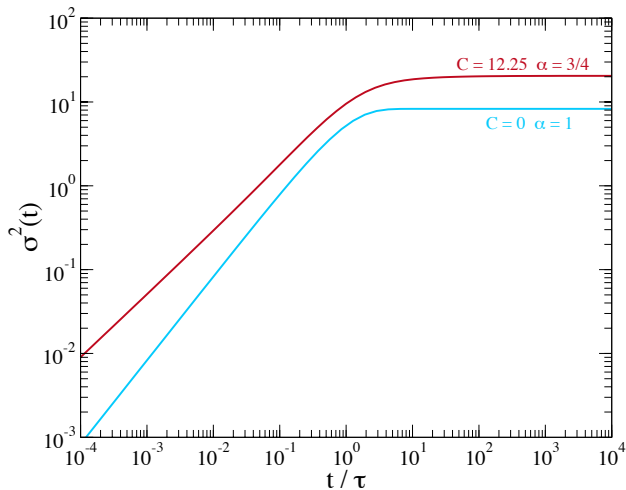}
\caption{Mean square distance travelled by a particle in a harmonic potential ($k=1$ pN/nm, $k_B T=4.14$ pN nm) in a viscous and non-active environment and in a viscoelastic active environment ($\alpha=3/4, \sqrt{C}= 3.5$ pN, $\tau_A=3$ s and $\tau=50 \mu$s.}
\label{Fig6}
\end{figure}

It requires further research to investigate whether the effects found here persists for more realistic models of the head linker and to quantify its impact on physical properties of the motor such as its velocity and processivity. 

\section{Particle in a double well} 
To understand how macromolecules fold into three-dimensional structures is one of the big open problems of biological physics. Folding reactions are often described in terms of a reaction coordinate $x$ which moves stochastically in a free energy landscape $V(x)$, i.e. whose time evolution is given by the Langevin equation (\ref{1}) \cite{Oliveberg05}. The reaction coordinate is a slow variable like the end-to-end distance of the polymer while $V(x)$ is obtained by averaging out the fast variables like the positions of individual monomers. Since this averaging cannot easily be performed exactly, it is often assumed that $V(x)$ has the shape of a double well whose minima correspond with the unfolded and folded molecule. 

One of the simplest and best understood folding processes is that of the zipping of a DNA or RNA hairpin. The folding of nucleic acids is simpler than that of proteins because the four nucleotides are chemically more similar than the 20 amino acids. Of all the possible secondary structures of nucleic acids, the hairpin is the simplest. In recent years, much advance has been made in measuring various properties of the zipping/unzipping of simple DNA and RNA structures, also at the single molecule level \cite{Borgia08,Woodside08}. In these latter experiments, the hairpins are held under tension by chemically connecting them to two beads that are in an optical trap. If the appropriate forces are applied, the free energy of the zipped and unzipped state are almost equal and the molecule continuously flips between the two states. From the measured dynamics, the free energy landscape for hairpins can then be determined. The folding or unfolding rate corresponds to the Kramers' rate for the particle to diffuse from one minimum to the other. Kramers' theory starts from the Langevin dynamics (\ref{1}) of a particle in a double well potential or its equivalent description using a Fokker-Planck equation \cite{VanKampen07}. Measured folding times are in the range from milliseconds to minutes, i.e. can be either slow or fast in comparison with active forces. 

As a simple model to investigate how crowding and nonequilibrium affect the zipped/unzipped transition for a hairpin under tension, we use again (\ref{5}) with the double well potential
\begin{eqnarray}
V(x) = \Delta U \left[\left(\frac{x}{b}-1\right)^2 -1 \right]^2
\label{28}
\end{eqnarray}
This potential has minima at $x=0$ and $2b$ and a maximum at $x=b$ of height $\Delta U$. It gives a reasonable description of the measured free energy landscapes of the hairpins 20TS06/T4 and 20TS10/T4 \cite{Neupane12}. 

Like in the case of the harmonic oscillator we can associate a typical time $\tau_w$ with the motion of a particle in the double well potential. $\tau_w$ is still given by (\ref{20}) where now $k$ is determined by making an harmonic approximation to (\ref{28}) around a minimum of the potential (i.e. $k=8 \Delta U/b^2$). Thus we define 
\begin{eqnarray}
\tau_w = \left(\frac {\eta_\alpha b^2}{8 \Delta U}\right)^{1/\alpha}
\label{29}
\end{eqnarray}
We also introduce the time $\tau_e=2^{1/\alpha} \tau_w$ associated with the top of the potential barrier. 

Using the results of \cite{Neupane12} we can estimate that for the DNA hairpins mentionned above, $\Delta U \approx 20$ kJ/mol ($=33$ pN nm) and $b\approx 7.5$ nm, from which it follows that $\tau_w \approx 8.8 \mu$s so that also in this case the biophysically interesting situation is the one where $\tau_w \ll \tau_A$.

Since the Langevin equation (\ref{5}) cannot be solved exactly for the potential (\ref{28}) we have developed an approach to numerically integrate Langevin equations with correlated noises and friction with a memory. It combines a number of existing algorithms such as the Hosking algorithm to generate fractional Gaussian noise. A full description of our method is given in the supplemental material. There we also compare the results of a simulation of a particle in a harmonic potential with the exact results of the preceding section. The fact that both approaches give the same results gives good confidence that our numerical technique works well. 

Using this numerical method, we have determined the probability distribution $P(x,t)$ that the particle in the double well is at position $x$ at time $t$. A typical result is shown in Fig. 7a. In this simulation the particles where all started at $x=b$. This corresponds to a nonequilibrium situation which allows us to clearly see the time evolution of $P(x,t)$ also in cases where the stationary distribution reached after a long time is the equilibrium one. 

We expect that for times $t \gg \tau_w$, $P(x,t)$ reaches a stationary distribution $P_{ss}(x)$, as can indeed be seen in Fig. 7a. In absence of active forces that stationary state should, {\it independently of $\alpha$},  correspond to the equilibrium one $P_{eq}(x,t)$ given by the Boltzmann distribution,
\begin{eqnarray}
P_{eq}(x,t) = \frac{1}{Z} \exp{\left(- V(x)/k_B T\right)}
\label{30}
\end{eqnarray}
where $Z$ is the partition function. Figure 7b shows that indeed this equilibrium distribution is reached for three different values of $\alpha$. This is another proof that our numerical approach is able to handle the dynamics in a viscoelastic environment correctly.

More interesting is of course the steady state that is reached in the presence of active forces. Some results are given in Fig. 7c and 7d. They give $P_{ss}(x)$ for three different values of $\alpha$ with $\tau_A \ll \tau_w$ and $\tau_A \gg \tau_w$ respectively. We see that, as was the case for the harmonic oscillator, the steady state distribution depends on $\alpha$ if $\tau_A \ll \tau_w$ (Fig. 7c). However, in the limit $\tau_A \gg \tau_w$ (Fig. 7d) this distribution becomes again independent of $\alpha$. This strongly suggests that in that limit the distribution can be described in terms of an equilibrium like distribution, albeit with a different potential $V_{eff}(x)$. 
%In fact, as was the case for the harmonic oscillator, once $\tau_A$ becomes greater than $\tau$ its actual value has little influence on $P_{ss}(x)$ ({\color{red} this has to be checked})

\begin{figure}[h]
\centering
\includegraphics[width=9cm]{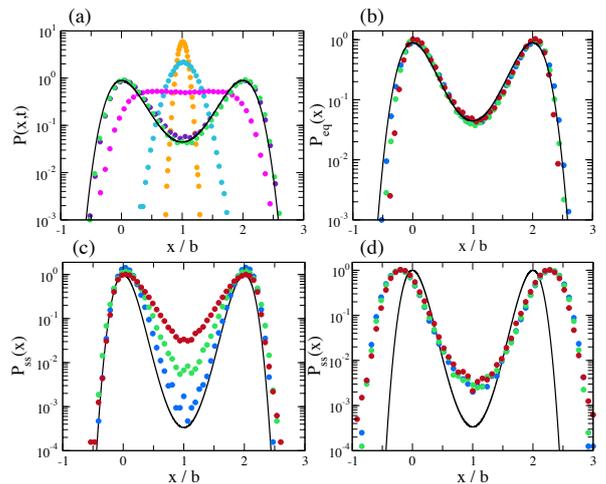}
\caption{In all panels, the black line represents the Boltzmann distribution \eqref{30} and the simulated data is presented with coloured dots. Panel (a) shows the probability distribution $P(x,t)$ for a particle in the double well surrounded by a viscoelastic bath at various times: $t/\tau_e=10^{-2}$ (orange), $10^{-1}$ (blue), $1$ (pink), $10$ (purple) and $10^{2}$ (green). The initial condition is $P(x,0)=\delta(x-b)$, while $\alpha=0.8$, $b=1$, $\Delta U=3$ and $k_BT=\gamma=1$. Panel (b) gives the equilibrium distribution $P_{eq}(x)$ for different values of $\alpha$ at $t/\tau_e=10^{2}$, with $\alpha=0.4$ (red), $0.6$ (blue) and $0.8$ (green). The other parameters are the same as in panel (a). Panels (c) and (d) show the stationary distribution $P_{ss}(x)$ at $t/\tau_e=10^{2}$ for a particle in a double well ($\Delta U=8$, $b=1$) surrounded by a viscoelastic bath ($\alpha=0.4$ (red), $0.6$ (blue), $0.8$ (green), $k_BT=\gamma=1$) in the presence of active forces with $C=10^3$ and short persistence time ($\tau_A/\tau_e=10^{-3}$) in panel (c) and large persistence time ($\tau_A/\tau_e=10^{3}$) in panel (d). }
\label{Fig7}
\end{figure}

This effective potential can, up to an additive constant, be obtained from $P_{ss}(x)$ as $V_{eff}(x) = -k_B T \ln P_{ss}(x)$. The result of applying this transformation to the results of Fig 7d is given in Fig. 8. Surprisingly, for the parameter values used ($C=10^3, \tau_A/\tau_e=10^3$), the effective potential can again be well fitted by the quartic potential (\ref{28}) where however the distance between the two minima, $b'$, has increased and the height of the potential barrier $\Delta U'$ is decreased. We estimate $b'/b= 1.27$ and $\Delta U'/\Delta U=0.72$. It would be interesting to develop an analytical approach that allows one to determine $V_{eff}(x)$ from $V(x)$ as was recently done for other problems in the field of active matter \cite{Fodor16}. 

\begin{figure}[h]
\centering
\includegraphics[width=9cm]{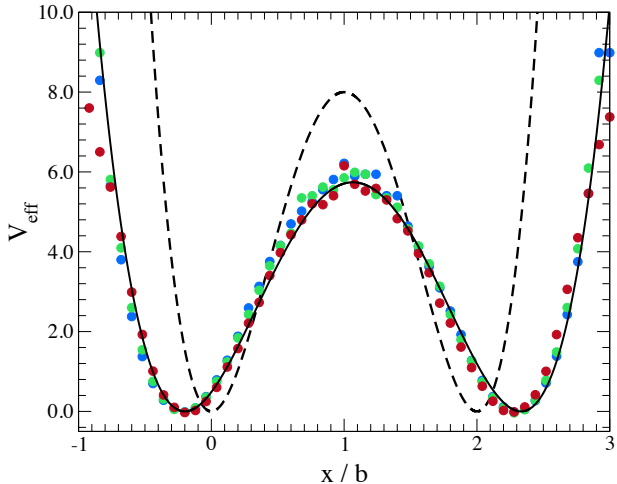}
\caption{The effective potential $V_{eff}(x)$ (see text) for the same data as shown in Fig. 7d. The dashed line shows the original potential $V(x)$ with $b=1$ and $\Delta U=8$. The full line is a fit through the simulation data with the potential (\ref{28}) but with modified parameters $b'=1.27$ and $\Delta U'=5.74$.}
\label{Fig8}
\end{figure}

We have also calculated the survival probability $Q(t)$ for a particle in one of the potential minima. We start with a particle in the left well (its position is drawn from the equilibrium distribution in an harmonic approximation of the left well) and determine the time it needs to reach the bottom of the right well, which acts as an absorbing boundary. From a large number of such simulations we can determine the survival probability $Q(t)$, i.e. the probability that the particle is still in the left well at time $t$. In the ordinary Kramers' problem, $Q(t)$ decays exponentially for $t$ large. The decay rate $\lambda_d$ is then given by Kramers' famous formula \cite{VanKampen07}
\begin{eqnarray}
\lambda_d =  \left( \frac{\left( V''(0) | V''(b)|\right)^{1/2}}{2 \pi \gamma} \right) e^{-\Delta U/k_B T}
\label{31}
\end{eqnarray}
In Fig. 9, we show some results of simulations of the survival probability in the regime $\tau_A \gg \tau_e$. One might be tempted to use Kramers' formula (\ref{31}) with the parameters of the effective potential, but this turns out not to work. As the inset of Fig. 9 shows the survival probability depends on $\alpha$ even though the stationary state distribution does not. The main part of the graph shows $Q(t)$ at fixed $\alpha=0.8$ for different values of the barrier height $\Delta U$. We observe a strange crossover phenomenon in which initially the particles escapes more rapidly for small barriers (as would be the case in equilibrium), but for larger times this trend is reversed. For these parameter values, the decay of $Q(t)$ is exponential for large enough times so that a unique transition rate can be defined. 

\begin{figure}[h]
\centering
\includegraphics[width=9cm]{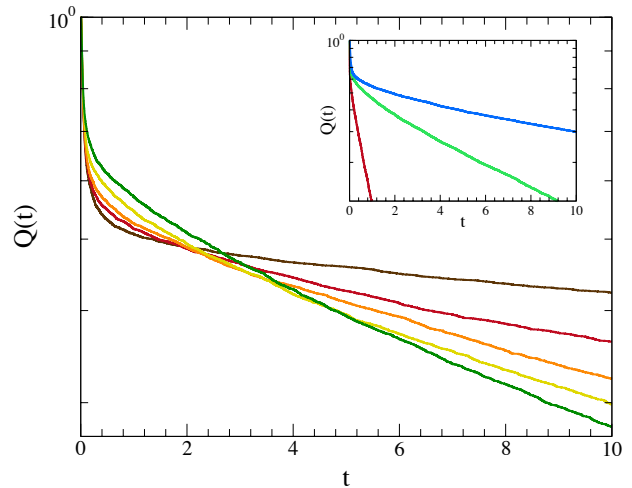}
\caption{Survival probability $Q(t)$ in the double well potential (\ref{28}) for $b=1$ for various values of $\Delta U$. The particle moves in a viscoelastic medium with $\alpha=0.8$ and is subject to active forces with $C=10^3,\tau_A/\tau_e=10^3$. The inset shows $Q(t)$ for $\Delta U=8$ and in media with $\alpha=0.8, 0.6$ and $0.4$ (top to bottom). The parameters of the active forces are the same as in the main figure.}
\label{Fig9}
\end{figure}

We have also performed similar simulations for the parameter values of the DNA-hairpins mentioned above.
Furthermore, we take the values $\alpha=0.75$ (since we are in the regime $\tau_A \gg \tau_w$, we only considered one value for $\alpha$) and $C=10$ pN$^2$, $\tau_A=3$ s. As can be seen in Fig. 10, the effective potential still has two minima. For values of $C$ that are not too big, the effective potential can again be fitted by (\ref{28}). We notice that the active fluctuations have greatly lowered the effective barrier between the two minima ($\Delta U'/\Delta U =0.23$) and at the same time has slightly increased the distance between the two minima ($b'/b=1.03$). We can expect that the active forces therefore will considerably lower the zipping/unzipping time of an DNA-hairpin since the escape time from one of the minima of the effective potential will be considerably lower. In order to quantify this result, we have numerically determined the survival probability in the potential well (\ref{28}) both for the viscous passive case, and for the viscoelastic active one. Results are shown in Fig. 11. We see that within the simulated time window, a particle in a viscous and passive environment (upper curve) is still almost surely in the left well whereas a particle in a viscoelastic and active environment has left that well with a probability $\approx 0.42$. These results indicate that the zipping/unzipping time can be largely reduced in an active environment. Moreover, within the times simulated, the survival probability doesn't decay exponentially, a phenomenon that was also found in other studies of escape times in viscoelastic, but non active, environments \cite{Goychuk07,Goychuk09}.

\begin{figure}[h]
\centering
\includegraphics[width=9cm]{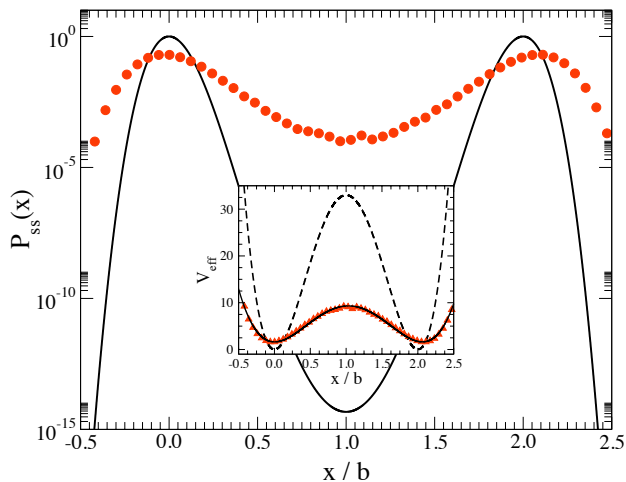}
\caption{Probability distribution $P_{ss}(x)$ (main figure)  and effective potential $V_{eff}(x)$ (inset) for the experimentally determined potential of DNA hairpins \cite{Neupane12}. The full line in the main figure shows the equilibrium distribution $P_{eq}(x)$ associated with this potential (with $b=7.5$ nm and $\Delta U=33$ pN nm - see also dashed line in inset). The full line in the inset is a fit with the quartic potential (\ref{28}) (with $b'=7.72$ nm and $\Delta U'=7.6$ pN nm) through the simulation data. }
\label{Fig10}
\end{figure}

\begin{figure}[h]
\centering
\includegraphics[width=9cm]{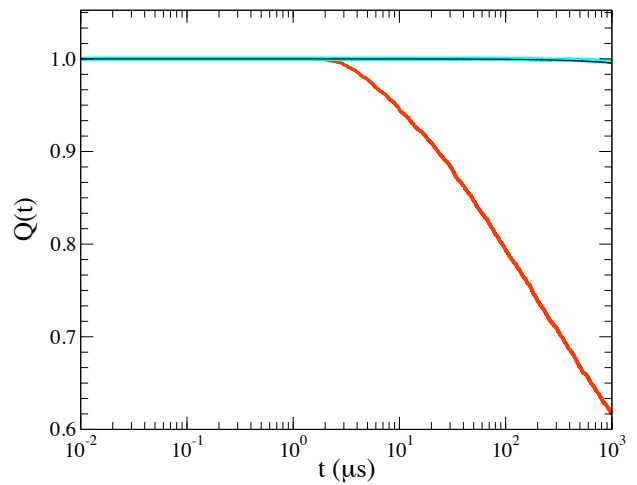}
\caption{Survival probability  in the double well potential (\ref{28}) for the experimentally determined potential of DNA hairpins \cite{Neupane12} in a viscous medium and in absence of active forces (blue curve) and in a viscoelastic environment ($\alpha=3/4$) in the presence of active forces $C=10$ pN$^2$ $,\tau_A=3$ s (orange curve). The black line through the upper curve is an exponential function decaying with the Kramers' rate (\ref{31}).}
\label{Fig11}
\end{figure}

\section{Discussion}
In this paper we have introduced a (generalized) Langevin equation for a particle in a viscoelastic and active environment. The latter gives rise to random forces on the particle. These forces are assumed to be centered Gaussian random variables with exponentially decaying autocorrelation. Their effect is to put the particle out of equilibrium. We have argued that our model can be used as a first step towards understanding various processes occuring in artificial cytoskeletons and in real cells. 
It is interesting to remark here that noise with a correlation (\ref{4}) has recently also been introduced in the study of active particles \cite{Farage15,Koumakis14,Szamel15,Fodor16}. These type of active particles have been named Ornstein-Uhlenbeck active particles (OUAP). Within this interpretation, Eq. (\ref{5}) can also be used to study the motion of OUAP in an external potential and in a viscoelastic environment. 

For the motion of free particles we have shown that the mean squared displacement shows a regime of superdiffusive motion (if at least $\alpha > 1/2$) followed by a subdiffusive one as observed in several recent experiments. We have also shown that the effective exponents measured do not only depend on the rheological properties of the environment but also on the characteristics of the active forces. On the one hand, this is a warning that care has to be taken in order to determine the "bare" value of $\alpha$ from experiments. On the other hand, it shows that  both $\alpha$, and the properties of the active forces (their strength $C$ and persistence time $\tau_A$) can be determined from measured data. 

For a particle in an harmonic potential, we have calculated  the time evolution of both the mean squared displacement and the variance of its position. We have shown that these quantities evolve to non-equilibrium steady state values which depend on $\alpha$. However, in the biophysically relevant regime where $\tau_A$ is large in comparison to the typical relaxation time in the harmonic well, this dependence disappears. We have applied our results to the motion of the tethered head of kinesin-1. Here we have found that active forces enlarge the probability that the head is near its binding site. Moreover, the active forces help in preventing the expected slow down of the motor in a crowded and viscoelastic environment. The latter conclusions need to be verified in more realistic models of the neck linker potential, for example by replacing the harmonic force on the particle by that of a wormlike chain. This can easily be done using our numerical approach to solving (\ref{5}). Of course, the stepping of the tethered head is only one step in the sequence of kinetic steps that describe the motion of kinesin \cite{Clancy11} and it therefore remains to be investigated how the influence of active forces on that step can modify the speed or processivity of kinesin. 

Finally we have investigated the motion of a particle in a double well potential. Using a numerical approach we determined the probability distribution $P(x,t)$ of the position of the particle and the survival probability $Q(t)$ of a particle starting in the left well. We have shown, that as was the case for the harmonic oscillator, $P(x,t)$ evolves towards a non-equilibrium steady state that depends on $\alpha$, but in the limit of large persistence times of the active forces that dependence disappears. Nevertheless, the survival probability is still dependent on this parameter. For different $\alpha$ we thus have the same probability that the particle is in one of the wells while the fluxes between the wells are different and increase with decreasing values of $\alpha$. We have applied this model to the zipping/unzipping of a DNA-hairpin held under tension in an optical trap. We have found that the active forces effectively lower the potential barrier between the two states, which lead to an escape rate that is higher than that in equilibrium. 
Clearly, the behaviour of a particle in a double well and in an active viscoelastic environment is very rich and deserves further investigation. Also, our observation that zipping/unzipping times can be decreased considerably needs further research. The literature on Kramers' theory is vast \cite{Haenggi90} and generalisations of Kramers' theory to viscoelastic environments have been discussed by various authors, without reaching a clear consensus \cite{Goychuk09}. We leave it to further work to develop a more complete theory of escape rates in environments that are both viscoelastic and out of equilibrium. Also simulations of zipping under tension of simple polymer models would be welcome in order to see whether the results found here in a description based on a reaction coordinate remain valid if all degrees of freedom of the polymer are taken into account.

\ \\
\ \\
{\bf Acknowledgement} The computational resources and services used in this work were provided by the VSC (Flemish Supercomputer Center), funded by the Research Foundation - Flanders (FWO) and the Flemish Government - department EWI. 
%%%REFERENCES%%%
 

\begin{thebibliography}{99}
\bibitem{remark}
The MSD of the free particle can be found by taking the limit $k \to 0$, i.e. $\tau \to \infty$ in (\ref{27}). Using the series expansion of the Mittag-Leffler function (\ref{17}) and performing one of the integrals appearing in the definition of $A_\alpha(t,\tau,\tau_A)$ then leads to (\ref{12}).
\end{thebibliography}
\end{document}